\documentclass[iop,apj,tighten]{emulateapj}
\usepackage{amsmath,amstext}
\usepackage{apjfonts}
\usepackage[breaklinks,colorlinks,citecolor=blue,linkcolor=magenta]{hyperref} 
\usepackage[all]{hypcap} 
\usepackage{float}
\usepackage{footmisc}
\usepackage{multirow}
\usepackage{lineno}

\def \nustar {{\em NuSTAR}}
\def \xmm {{\em XMM-Newton}}

\def \swift {{\em Swift}}

\def \swiftsource {{Swift J1658.2--4242}}

\shortauthors{Xu et al.}

\begin{document}

\title{Broadband X-ray Spectral and Timing  Analyses of the Black Hole Binary Candidate \\ Swift J1658.2-4242: Rapid Flux Variation and the Turn-on of a Transient QPO}

\author{Yanjun Xu\altaffilmark{1}}
\author{Fiona A. Harrison\altaffilmark{1}}
\author{John A. Tomsick\altaffilmark{2}}
\author{Didier Barret\altaffilmark{3,4}}
\author{Poshak Gandhi\altaffilmark{5}}
\author{Javier A. Garc\'ia\altaffilmark{1,6}}
\author{Jon M. Miller\altaffilmark{7}}
\author{Phil Uttley\altaffilmark{8}}
\author{Dominic J. Walton\altaffilmark{9}}

\affil{$^{1}$ Cahill Center for Astronomy and Astrophysics, California Institute of Technology, Pasadena, CA 91125, USA}
\affil{$^{2}$ Space Sciences Laboratory, 7 Gauss Way, University of California, Berkeley, CA 94720-7450, USA}
\affil{$^{3}$ Universite de Toulouse, UPS-OMP, IRAP, Toulouse, France}
\affil{$^{4}$ CNRS, IRAP, 9 Av. colonel Roche, BP 44346, F-31028 Toulouse cedex 4, France}
\affil{$^{5}$ Department of Physics and Astronomy, University of Southampton, SO17 3RT, UK}
\affil{$^{6}$ Remeis Observatory \& ECAP, Universit\"at Erlangen-N\"urnberg, Sternwartstr.~7, 96049 Bamberg, Germany}
\affil{$^{7}$ Department of Astronomy, University of Michigan, 1085 South University Avenue, Ann Arbor, MI 48109, USA}
\affil{$^{8}$ Anton Pannekoek Institute, University of Amsterdam, Science Park 904, 1098 XH Amsterdam, The Netherlands}
\affil{$^{9}$ Institute of Astronomy, University of Cambridge, Madingley Road, Cambridge CB3 0HA, UK}

\begin{abstract}
We report results from joint \nustar, \swift\ and \xmm\ observations of the newly discovered black hole X-ray binary candidate \swiftsource\ in the intermediate state. We observe a peculiar event in this source, with its X-ray flux rapidly decreasing by $\sim$45\% in $\sim$40~s, accompanied by only subtle changes in the shape of the broadband X-ray spectrum. In addition, we find a sudden turn-on of a transient QPO with a frequency of $6-7$~Hz around the time of the flux change, and the total fractional rms amplitude of the power spectrum increases from $\sim$2\% to $\sim$10\%. X-ray spectral and timing analyses indicate that the flux decrease is driven by intrinsic changes in the accretion flow around the black hole, rather than intervening material along the line of sight. In addition, we do not significantly detect any relativistic disk reflection component, indicating it is much weaker than previously observed while the source was in the bright hard state. We propose accretion disk instabilities triggered at a large disk radius as the origin of the fast transition in spectral and timing properties, and discuss possible causes of the unusual properties observed in \swiftsource. The prompt flux variation detected along with the emergence of a QPO makes the event an interesting case for investigating QPO mechanisms in black hole X-ray binaries.    

\end{abstract}

\maketitle

\section{INTRODUCTION}
Most galactic black hole X-ray binaries are discovered as transients, when they go into bright outbursts in the X-ray band. During typical outbursts, these systems are known to undergo undergo a transition from the low/hard to the high/soft state through relatively short-lived intermediate states, with these states exhibiting distinct spectral and timing properties \citep[see][for reviews]{rem06,belloni2016}. \swiftsource\ is a new black hole X-ray binary candidate, first reported by \swift-BAT on 2018 February 16 \citep{bart18}. Subsequent observations exhibiting dips in the light curve indicate that it is highly absorbed and is viewed at a high inclination angle \citep{beri18, lien18,xu_1658}. Relativistic disk reflection features, including a broad asymmetric Fe K$\alpha$ line peaking at 6--7~keV and Compton reflection hump around 30~keV, were detected by \nustar\ when \swiftsource\ was in the low/hard state \citep{xu_1658}. Relativistic reflection features in the X-ray spectrum of Galactic black hole binaries are believed to arise from reprocessing of hard X-ray emission from the corona by the inner accretion disk \citep[][]{fabian89, reynolds14}. Detailed modeling of the reflection spectra of \swiftsource\ suggests that the central object in the system is a rapidly spinning black hole ($a^*>0.96$), and also supported the conclusion that we view the system at high inclination, finding $\theta=64^{+2 \circ}_{-3}$ for the inner disc \citep{xu_1658}. 

Variability on a wide range of timescales is characteristic of the accretion process around both stellar-mass and supermassive black holes \citep[e.g.,][]{ulrich97, vander06, mchardy10}. Rapid flux variations are commonly found in the X-ray light curves of black hole binaries, and are believed to probe the properties of the inner accretion flow in the vicinity of stellar-mass black holes. The power spectrum generated from the X-ray light curve of a black hole X-ray binary is typically characterized by a variable broadband noise component along with transient and discrete peaks superimposed on top of the continuum \citep[e.g.,][]{van89, vander06}. The peaks are termed quasi-periodic oscillations (QPOs). Low-frequency QPOs observed in black hole binary systems, with typical frequencies from a few mHz to 30~Hz, can be generally classified into three main types (type-A, B and C), based on their characteristics in the power density spectrum \citep[see][for details]{wij99, cas04, cas05}. Type-A QPOs are characterized by a weak and broad peak in the power spectrum. The most common type of low-frequency QPOs are type-C QPOs, which are usually strong and span the frequency range of $\sim$0.1--30~Hz. Type-B QPOs are rarer, limited to a narrow frequency range (typically $\sim$5-6~Hz), and can be distinguished by the fractional amplitude and underlying noise shape. Type-B QPOs are usually observed during the time of transitions between hard-intermediate (HIMS) and soft-intermediate (SIMS) states (or transitions between hard and soft states), and have been proposed to be associated with major jet ejection events \citep[e.g.,][]{sole08,fender09}. There is evidence that type-B QPOs are stronger in systems that are viewed close to face-on, while type-C QPOs are stronger in high-inclination systems \citep[][]{motta15}, supporting the idea that the two types have different physical origins. The rapid transitions of QPOs have been found in a few black hole binaries or black hole candidates via time-resolved analysis: {GS 1124--684} \citep{taki97}, {XTE J1859+226} \citep{cas04, bell05}, {GX 339--4} \citep{nes03}, {XTE J1550--564} \citep{srir13}, {XTE J1859+226} \citep{srir16}, and also recently in {MAXI J1535--571} \citep{huang18}, in most cases type-B QPOs are identified. 

Despite the richness in observational phenomenology, the physical origins of QPOs are still highly uncertain. For low-frequency QPOs, the frequency is much longer than the dynamical timescales in the strong gravity regime of stellar-mass black holes. Different theoretical models have been put forward to provide a physical explanation for low-frequency QPOs: some models invoke geometric effects, e.g., Lense-Thirring precession of a radially extended section of the inner accretion flow \citep[e.g,][]{stellar99, ingram09, motta18}, which is a General-Relativity frame-dragging effect; other models consider intrinsic instabilities in the accretion flow, e.g., oscillations caused by standing shocks \citep{cha93} or magneto-acoustic wave propagation \citep{tita04,cabanac10}. Similarly, various attempts have also been made to understand the broadband aperiodic noise continuum in the power spectrum \citep[e.g.,][]{lyub97,arevalo06,ingram10}, with the propagation of fluctuating accretion being a popular model, however there is still no consensus on the origin. X-ray spectral-timing studies are believed to be a promising method to investigate the nature of QPOs.

In this work, we present X-ray spectral and timing analyses of the new black hole binary candidate \swiftsource\ in the intermediate state of its 2018 outburst, observed by \nustar, \swift\ and \xmm. The paper is structured as follows: in Section~\ref{sec:sec2}, we describe details of the observations used in this work and the data reduction procedures; results from our X-ray timing and spectral analyses are present in Section \ref{sec:sec3} and Section \ref{sec:sec4}, respectively; finally, we discuss the physical implications from the observations in Section \ref{sec:sec5}, and summarize the paper in Section \ref{sec:sec6}.

\section{Observations and DATA REDUCTION} 
\begin{figure*}
\centering
\includegraphics[width=0.95\textwidth]{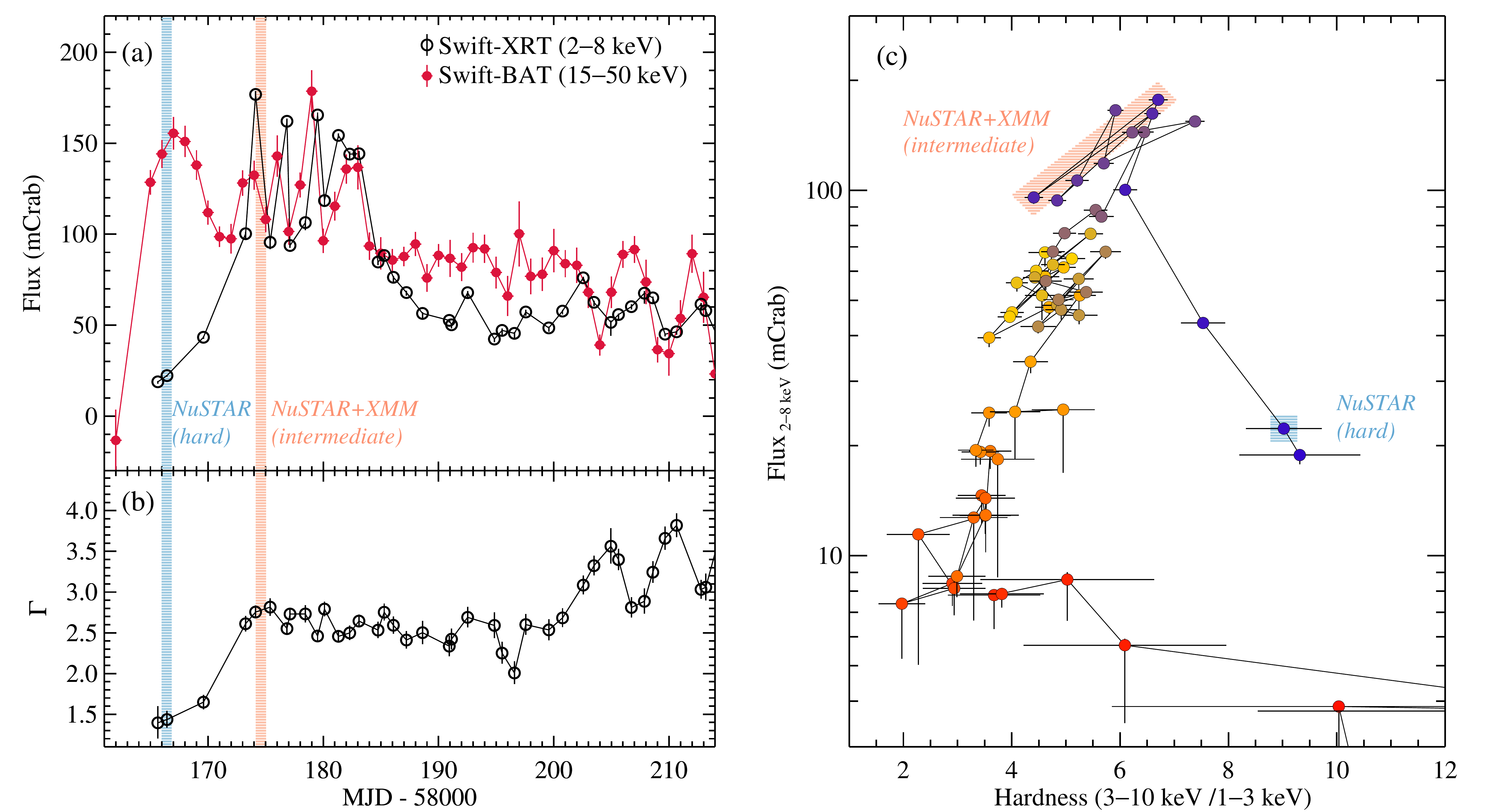}
\caption{(a) (b) Long term \swift-XRT and \swift-BAT light curves of \swiftsource\ since the beginning of its 2018 outburst. The BAT light  curve in daily averaged flux is from the \swift-BAT transient monitor \citep{swiftbat}, and is rescaled to the unit mCrab from count rates in the 15--50 keV band. The XRT flux was calculated by fitting the spectra with an absorbed power-law model. (c) HID calculated with \swift-XRT data. The color scheme indicates the time since the start of the observation. Hardness is calculated by the count rate ratio between 3--10 keV and 1--3 keV. The blue shaded area marks the first \nustar\ observation, which caught the source in the rising hard state \citep[see detailed data analysis in][]{xu_1658}. Our joint \xmm\ and \nustar\ observations used in this work is marked in orange, which caught the source in the intermediate state, transitioning from a canonical hard to a canonical soft state.
\label{fig:fig1}}
\end{figure*}

{\em{Nuclear Spectroscopic Telescope Array}} (\nustar; \citealt{nustar}) and \xmm\ \citep{xmm} jointly observed \swiftsource\ on 2018 February 25 (MJD 58175). The outburst of \swiftsource\ was also monitored daily by the X-ray Telescope (XRT; \citealt{swiftxrt}) on the Neil Gehrels \swift\ Observatory (\swift; \citealt{swift}). The joint \nustar\ and \xmm\ observations caught \swiftsource\ shortly after significant spectral softening occurred during the rising phase of outburst, close to the time of the hard-to-soft state transition (see \swift\ monitoring light curves in Figure \ref{fig:fig1}(a)). Based on the hardness-intensity diagram (HID, Figure~\ref{fig:fig1}(c)), \swiftsource\ is consistent with being in an intermediate state, transitioning from a canonical hard to a canonical soft state, at a phase of an outburst when some black hole binaries are known to display repeated fast transitions back and forth involving HIMS and SIMS \citep[e.g.,][]{belloni2016}.

\label{sec:sec2}
\subsection{NuSTAR}
We reduced the \nustar\ data of \swiftsource\ (OBSID: 80301301002) using NuSTARDAS pipeline v.1.6.0 and CALDB v20180419. After standard data filtering with NUPIPELINE, the exposure times are 31.5~ks and 31.8~ks for the two focal plane modules, FPMA and FPMB, respectively. The source spectra and light curves were extracted from a circular region with the radius of 180$\arcsec$ centered on the location of \swiftsource\ using NUPRODUCTS. Background was estimated from source-free areas using polygonal regions. The \nustar\ spectra were grouped to have a signal-to-noise ratio (S/N) of at least 30 per bin.

\subsection{XMM-Newton}
The \xmm\ observation of \swiftsource\ (OBSID: 0802300201) were processed using the \xmm\ Science Analysis System (SAS) v17.0.0 following standard procedures. The EPIC-MOS1 and EPIC-MOS2 cameras were closed during the observation. The prime instrument we use is EPIC-pn \citep{epic_pn}. The first part of the observation was carried out in the EPIC-pn Timing Mode, and was later switched to the Burst Mode (a special flavour of the Timing Mode with low live time) due to the unanticipated high count rate. We selected events with Pattern $\leqslant$ 4 (singles and doubles) and Quality Flag = 0. The source region was chosen as rows 20 $\leqslant$  RAWX $\leqslant$ 55 (Timing Mode), 25 $\leqslant$ RAWX $\leqslant$ 50 and RAWY $\leqslant$ 142 (Burst Mode). Corresponding background was extracted from rows 3 $\leqslant$ RAWX $\leqslant$ 6. The resulting exposure times after data filtering are 41~ks for the Timing Mode, and 630~s for the Burst Mode. The source light curves were corrected for dead time, PSF and other instrumental effects using the {\tt epiclccorr} tool. Due to an apparent disagreement of \xmm\ data in spectral slope with \nustar\ and \swift-XRT, we do not use \xmm\ spectra for spectral analysis in this work. The spectral slope disagreement is likely to be associated with pile-up issues along with possible calibration uncertainties for bright targets observed by \xmm\ in the Timing Mode, which cannot be completely solved by excising data from the core of the \xmm\ PSF.

\begin{figure*}
\centering
\includegraphics[width=0.85\textwidth]{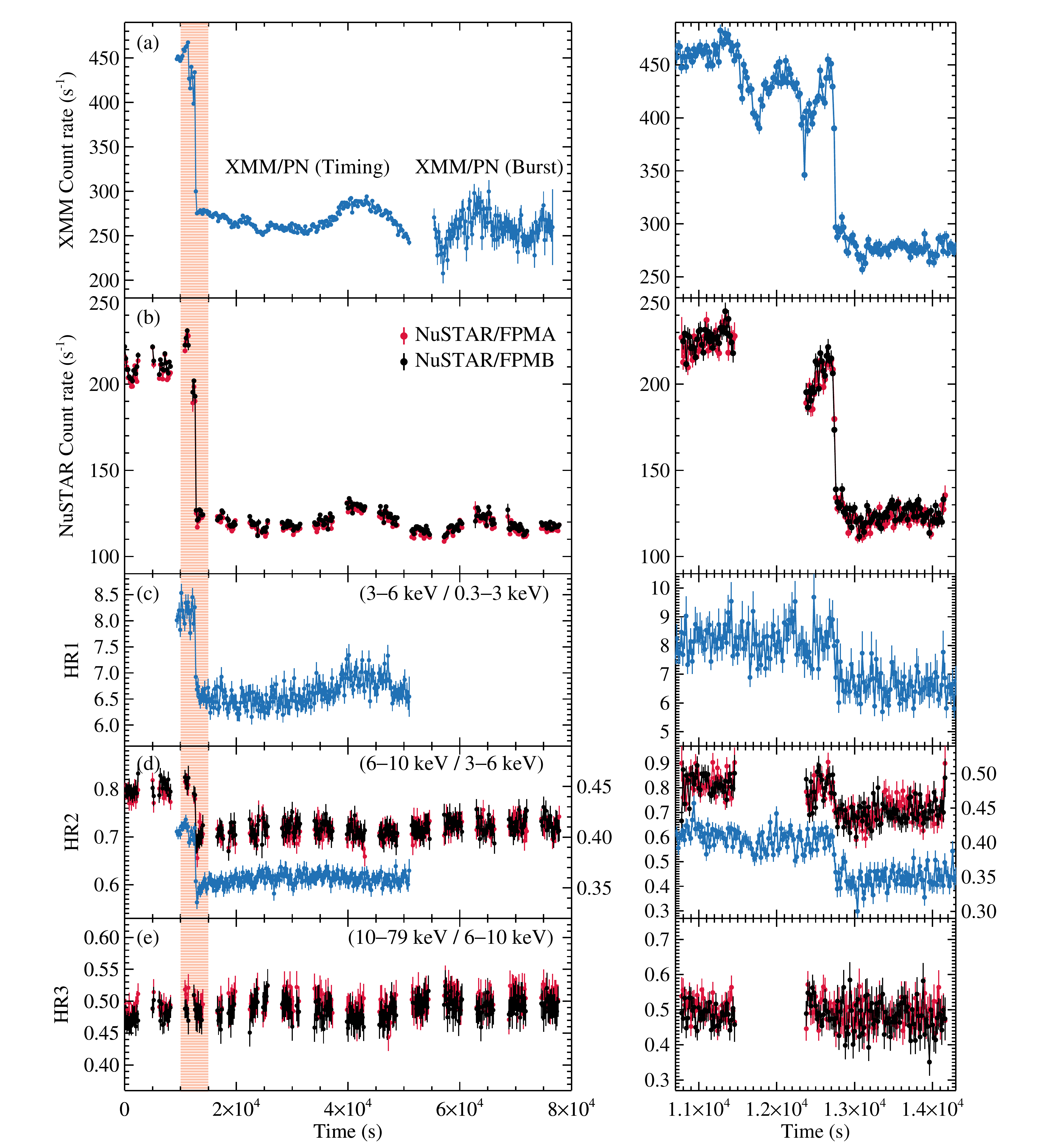}
\caption{(a) (b) X-ray light curves of \swiftsource\ from the joint \nustar\ and \xmm\ observations. A sharp drop in flux was observed by both telescopes around $1.27\times10^4$~s. Shaded area in the left panel marks the time interval around the flux drop, which is zoomed in for display clarity in the right panel. The left panel is plotted in 100~s bins, the time bins are 20~s for the right panel. \xmm\ EPIC-pn light curve (0.3--10~keV) is plotted in blue. \nustar\ FPMA and FPMB light curves (3--79~keV) are plotted in red and black, respectively. The gaps in the \nustar\ light curves are due to occultations and SAA passages. (c)--(e) Hardness ratios calculated from count rates in different energy bands. In panel (d), the left y-axis is the \nustar\ hardness ratio, and the right y-axis is the \xmm\ hardness ratio. The \xmm\ EPIC-pn Burst Mode data are not displayed in hardness ratios, as the the error bars are much larger than those of \nustar\ and \xmm\ Timing Mode, which is due to the low instrument live time of the Burst Mode.
\label{fig:fig2}}
\end{figure*}

\subsection{Swift}
Two \swift-XRT observations of \swiftsource\ (OBSID: 00010571004, 00010571005) were taken close to the time of the joint \nustar\ and \xmm\ observations, on 2018 February 25 and 26, respectively. The data were taken in the Windowed Timing mode. We extracted source spectra from a circular region with the radius of 70$\arcsec$, and the background was extracted from an annulus area with the inner and outer radii of 200$\arcsec$ and 300$\arcsec$. The averaged Swift-XRT count rates are $\sim$38 ct~s$^{-1}$ and $\sim$23 ct~s$^{-1}$ for the first and the second observation, respectively, thus the data are not affected by pile-up distortions. After standard data filtering, the exposure times are 857~s and 795~s for the first and second \swift-XRT observations. The \swift-XRT spectra were rebinned to have a S/N of least 5 per energy bin.

\section{Variability}
\label{sec:sec3}
\begin{figure*}
\centering
\includegraphics[width=0.75\textwidth]{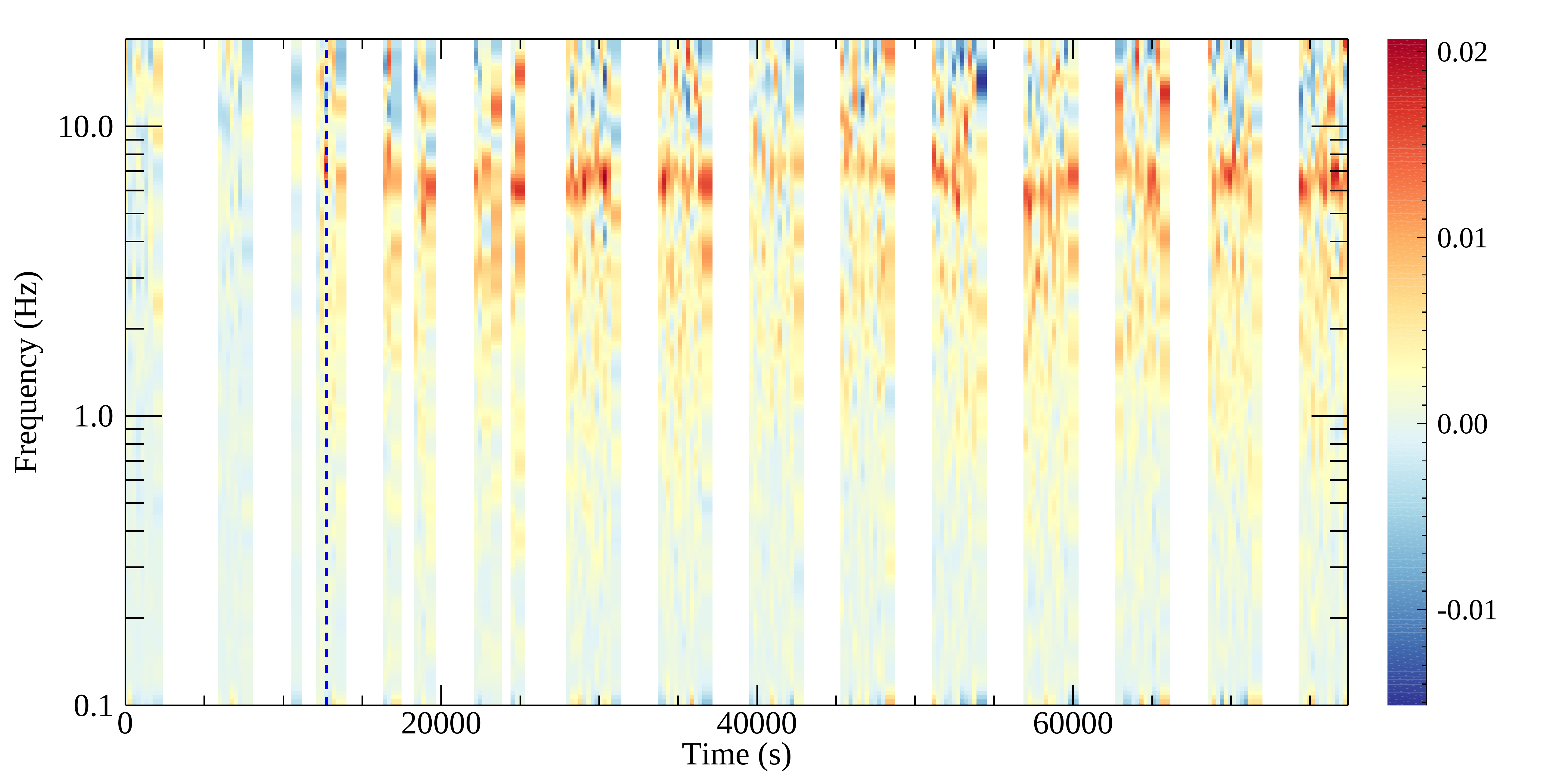}
\caption{Dynamical \nustar\ power spectrum in $\nu P_{\nu}$ representation, displaying the appearance of a QPO with the frequency of 6--7 Hz starting from $\sim$$1.3\times10^4$~s. The appearance of a QPO is found around the same time of the rapid flux variation in the X-ray light curves (marked by a blue dashed line).
\label{fig:fig3}}
\end{figure*}

\begin{figure}
\centering
\includegraphics[width=0.50\textwidth]{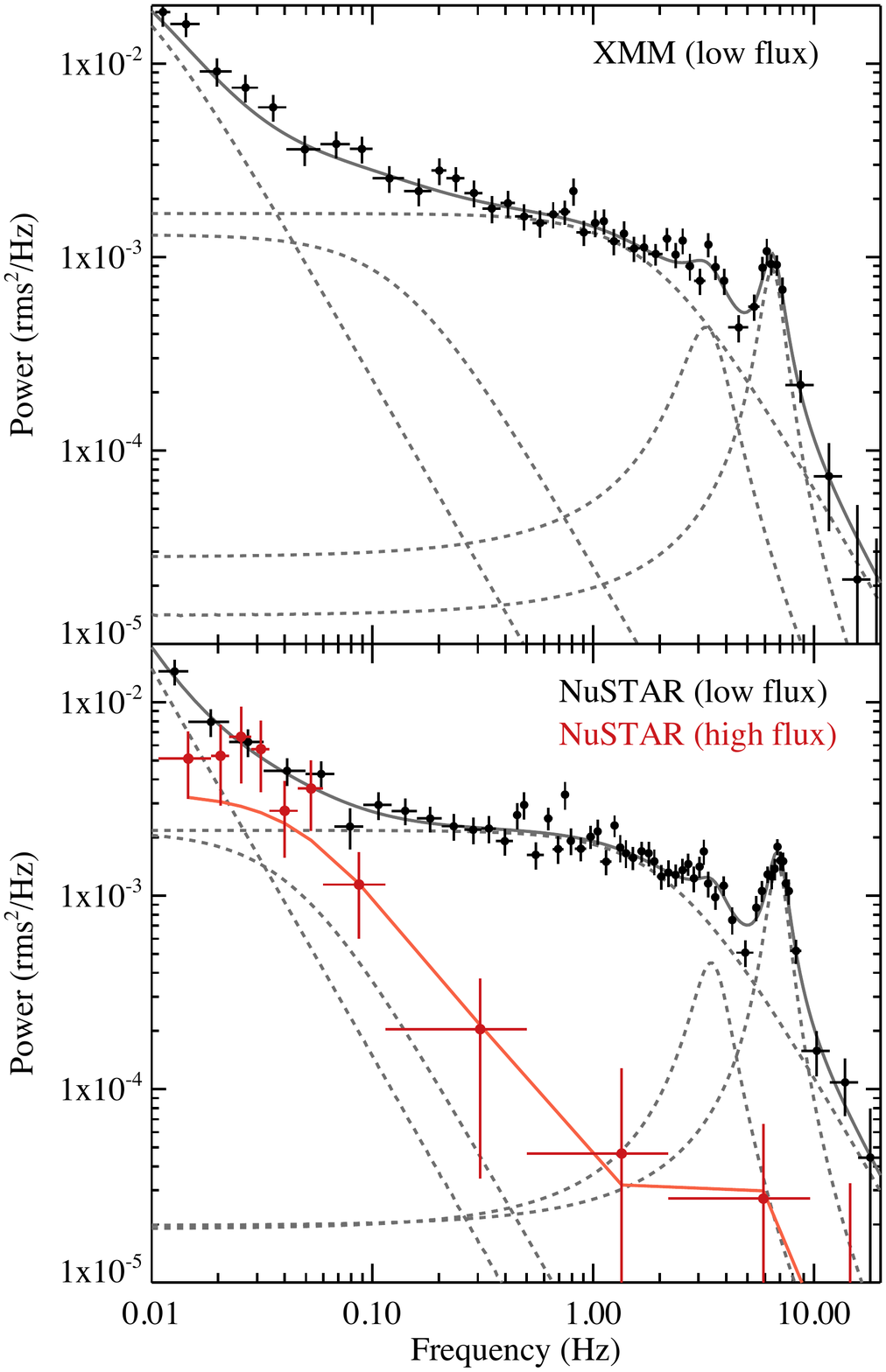}
\caption{Time-averaged power spectra of \swiftsource\ in the intermediate state with the best-fit model. The \nustar\ and \xmm\ power spectra are calculated in the energy band of 3--79~keV and 0.3--10~keV, respectively. The \nustar\ power spectra is plotted in black for the low-flux epoch, and in red for the high-flux epoch. The dashed lines mark the individual Lorentzian model components.
\label{fig:fig4}}
\end{figure}

A dramatic decrease in the flux of \swiftsource\ was simultaneously observed by \nustar\ and \xmm\ at 07:00:14 UTC on 2018 February 25 (around 1.27$\times10^4$~s in Figure~\ref{fig:fig2}). The source count rate decreased by $\sim$35\% in the \xmm\ band (0.3--10~keV), and $\sim$40\% in the \nustar\ band (3--79~keV) within only $\sim$40~s. The flux levels before and after this sudden flux variation were relatively stable, although increased variability can be seen in the \xmm\ light curve right before the large flux drop around $1.27 \times10^4$~s (see Figure~\ref{fig:fig2}, right panel). The timing of the sharp count rate drop is aligned very well in the \nustar\ and \xmm\ light curves, showing no noticeable energy-dependent time delay. We henceforth refer to the time intervals before and after the large flux drop as the high- and the low-flux epoch, respectively. The same trend of flux variation is also reflected in the long term \swift\ light curves of \swiftsource\ (Figure~\ref{fig:fig1}), where there is a flux difference of about 50\% between the two \swift-XRT exposures taken shortly before and after our joint \nustar\ and \xmm\ observations. 

Despite the dramatic change in flux level, there is only subtle variation in the broadband X-ray spectral energy distribution. We extracted source light curves taken by \nustar\ and \xmm\ in four different energy intervals, and calculated hardness ratios based on count rates in the corresponding energy bands, defined as: ${\rm HR1={C_{3-6~keV}/C_{0.3-3~keV}}}$, ${\rm HR2={C_{6-10~keV}/C_{3-6~keV}}}$ and ${\rm HR3={C_{10-79~keV}/C_{6-10~keV}}}$. The sudden drop in count rates around $1.27\times10^4$~s is evident in all four energy bands, and the changes observed in hardness ratios are small. As shown in Figure~\ref{fig:fig2}, HR1 and HR2 decrease by $\sim$20\% and $\sim$12\%, respectively, at the time of the flux decrease, while HR3 remains constant. The small decrease in hardness ratios below 10~keV (HR1 and HR2) indicates that the source energy spectrum turns slightly softer when the flux is lower, ruling out the possibility that the flux variation is caused by photoelectric absorption from material temporarily obscuring the line of sight. In addition, the constant hardness ratio (HR3) in the hard X-ray band throughout the observations suggests that the non-thermal spectral shape of \swiftsource\ remains roughly the same despite of the large flux change.

For X-ray timing analysis, we generated power spectra of \swiftsource\ from \nustar\ and \xmm\ data. We first applied barycenter corrections to the event files, transferring the photon arrival times to the barycenter of the Solar system using JPL Planetary Ephemeris DE-200. The \nustar\ power spectra were produced using MaLTPyNT in the energy band of 3--79~keV, with a light curve binning size of $2^{-6}$~s, and was averaged in 256~s intervals. The power spectrum generated by MaLTPyNT is the cross-power density spectrum of FPMA and FPMB, which helps to reduce deadtime distortions \citep{nustartiming}. We generated an \xmm\ EPIC-pn  power spectrum using the {\tt powspec} tool in the XRONOS package \citep{xronos}. The EPIC-pn power spectrum was calculated  in the 0.3--10~keV band with a light curve time resolution of 0.01~s, and was averaged from 8 intervals. Both power spectra were generated using the root-mean-square (rms) normalization, and were geometrically rebinned by a factor of 1.03 to reach nearly equally spaced frequency bins in the logarithmic scale.

The dynamical \nustar\ power spectrum is displayed in Figure~\ref{fig:fig3}. The first two orbits of data lack any significant periodicity. A QPO appears in the power spectrum at $\sim$$6-7$~Hz from $\sim$1.3$\times10^4$~s, coinciding with the time of the rapid flux decrease found in Figure~\ref{fig:fig2}. For further analysis, we generate \nustar\ and \xmm\ power density spectra for the high- and low-flux epochs separately.

We fit the time-averaged power spectra of \swiftsource\ during the two epochs with a unity response file in XSPEC v12.9.0n \citep{xspec}. In this work, we perform all spectral fitting in XSPEC using $\chi^2$ statistics, and all parameter uncertainties are reported at the 90\% confidence level for one parameter of interest unless otherwise clarified. The power density spectra of \swiftsource\ can be adequately fitted with a multi-Lorentzian model, which is commonly used for black hole binaries \citep{belloni02}. We use three Lorentzians to fit for the noise continuum, with the centroid frequency fixed at zero; one Lorentz function for the QPO, and one for the possible sub-harmonic with the frequency linked with half the fundamental QPO frequency. 

 Comparing the the time-averaged \nustar\ power spectra at the high- and low-flux epochs (see Figure~\ref{fig:fig4}), it is evident that the source variability increased significantly after the rapid flux drop: the total fractional rms variability (rms$_{\rm tot}$, integrated in the frequency range of 0.1--20~Hz) in the \nustar\ band increases from $2.4^{+1.0}_{-0.8}$\% to $10.8\pm0.2$\%, and a QPO peak emerges in the power spectra along with increased noise continuum above 0.1~Hz. 

\capstartfalse
\begin{deluxetable}{ccccccc}
\tablewidth{\columnwidth}
\tablecolumns{7}
\tabletypesize{\scriptsize}
\tablecaption{Power Spectra Properties \label{tab:tab1}}
\tablehead{
\colhead{}       
& \colhead{$\nu$$_{\rm qpo}$}                   
& \colhead{rms$_{\rm qpo}$}
& \colhead{$Q$}
& \colhead{rms$_{\rm tot}$}
& \colhead{$\chi^2/\nu$}\\
& \colhead{(\rm Hz)} 
& \colhead{($\%$)}
& \colhead{($\nu$/FWHM)}
& \colhead{($\%$)}
& \colhead{}
} 
\startdata
\multicolumn{6}{c}{High-flux Epoch} \\
\noalign{\smallskip}
\hline
\noalign{\smallskip} 
{\nustar}   &$6.83~(\rm fixed)$   & <2.4   &$3.9~(\rm fixed)$   &$2.4^{+1.1}_{-0.8}$ &$0.94$\\     
\noalign{\smallskip}          
\hline
\noalign{\smallskip} 
\multicolumn{6}{c}{Low-flux Epoch} \\
\noalign{\smallskip} 
\hline
\noalign{\smallskip} 
{\em XMM}      &$6.47\pm 0.15$  &$4.7\pm 0.4$     &$4.3\pm0.7$   &$9.0\pm0.2$ &$1.00$\\
\noalign{\smallskip} 
{\nustar}   &$6.83\pm 0.09$  &$5.9\pm 0.4$     &$3.9\pm0.7$ &$10.8\pm0.2$ &$1.05$                               
\enddata  
\tablecomments{
The timing properties are calculated in the energy band of $3-79$~keV with \nustar\ data, and in the band of $0.3-10.0$~keV with \xmm\ EPIC-pn Timing Mode data. The total fractional rms amplitude of the power spectra, rms$_{\rm tot}$, is integrated in the frequency range of 0.1--20~Hz. \xmm\ data in the high-flux epoch are not listed due to its limited exposure time. $\chi^2/\nu$ is the reduced chi-squared of the best-fit of the power spectra.}
                                                                                
\end{deluxetable}

The time-averaged QPO frequency measured at the low-flux epoch is $\nu_{\rm 3-79~keV}=6.83\pm0.09~{\rm Hz}$ in the \nustar\ band, and is $\nu_{\rm 0.3-10~keV}=6.47\pm0.15~{\rm Hz}$ in the \xmm\ band. The quality factor $Q~(\nu/{\rm FWHM})$ of the QPO is $4.3\pm0.7$ measured by \nustar, and is $3.9\pm0.7$ measured by \xmm, with the fractional rms amplitude, rms$_{\rm qpo}$, of $5.9\pm0.4\%$ and $4.7\pm0.4\%$, respectively (see details in Table~\ref{tab:tab1}). The characteristics of the QPO are similar to those of type-B or type-C low-frequency QPOs in black hole binaries \citep[e.g.,][]{cas04,cas05, motta15}. Type-C QPOs are usually strong, but can also be weak when appear in the soft state (fractional rms amplitude $\sim$1--25\%). They are variable in the frequency range of $\sim$0.1--30~Hz, and have a strong flat-top noise continuum in the power spectrum. Type-B QPOs are typically weaker (fractional rms amplitude $\sim$1--10\%), characterized by a weak red noise continuum, and are usually detected in a narrow frequency range around 5--6~Hz. We note that considering that the QPO peak lies on top of a strong noise continuum in the power spectra in Figure~\ref{fig:fig4}, the low-frequency QPO detected in \swiftsource\ is most likely to be a type-C QPO. The lack of a type-B QPO here is unusual, as a switch between a type-B QPO and a type-A/type-C QPO or noise is expected during fast transitions between HIMS and SIMS \citep[e.g.,][]{bell10,belloni2016}. The QPO frequency and strength increase at higher energies, as can been seen by comparing the QPO properties in the two different instrument bands, which is typical for QPOs observed in black hole binaries \citep[e.g.,][]{rod02,cas04}.

 The QPO is absent in the high-flux epoch. By using a Lorentzian function to fit for the possible presence of a QPO in the high-flux epoch, with the centroid and width of the Lorentzian fixed at the low-flux values, we can put an upper limit on the fractional rms amplitude of the QPO of < 2.4\% (90\% confidence level, in the 3--79 keV band of \nustar). We can rule out the possibility that the non-detection of QPO in the high-flux epoch is simply due to dilution by increased X-ray photons: the rms amplitude of the QPO in the low-flux epoch is 5.9\%, the value would be 3.5\% if diluted by extra 40\% of non-variable photons, which exceeds the upper limit obtained. Therefore, the non-detection of a QPO during the start of the observation when the count rate is high is because of the intrinsic weakness of the QPO signal.

\section{SPECTRAL ANALYSIS}
\label{sec:sec4}
For spectral modeling, we separately extracted \nustar\ energy spectra accumulated before and after the flux change. The \nustar\ exposure times during the high- and the low-flux epoch are $\sim$3.7~ks and $\sim$27.4~ks, respectively. We first fit the \nustar\ spectra in the two epochs jointly with a simple absorbed cutoff power-law model plus a thermal disk blackbody component, {\tt TBnew*(diskbb+cutoffpl)} (Model 1), in XSPEC notation. The {\tt cutoffpl} model is widely used to fit the non-thermal X-ray emission in black hole binaries. It is used as a phenomenological spectral description of the Comptonized emission generated in the corona in the vicinity of a black hole. Neutral absorption is accounted for by using the {\tt TBnew} absorption model, with the cross-sections from \cite{crosssec} and abundances from \cite{wil00}. In the case of \swiftsource, neutral absorption is mostly intrinsic to the source, as the absorption column density, $N_{\rm H}$, measured by modeling the X-ray spectrum, greatly exceeds the Galactic value of $N_{\rm H, Gal}=1.55\times10^{22}$~cm$^{-2}$ \citep{kal05}. All abundances are fixed at the Solar value in the {\tt TBnew} model. Changes in the shape of the non-thermal spectra of \swiftsource\ before and after the large flux variation are minimal, as revealed by the constant values of HR3 (see Figure~\ref{fig:fig2}(e)). Therefore, we link the parameters of the {\tt cutoffpl} model between epochs, including the power-law index, $\Gamma$, and the exponential high-energy cutoff, $E_{\rm cut}$, while allow the normalization of the {\tt cutoffpl} model to vary independently.

\begin{figure}
\centering
\includegraphics[width=0.48\textwidth]{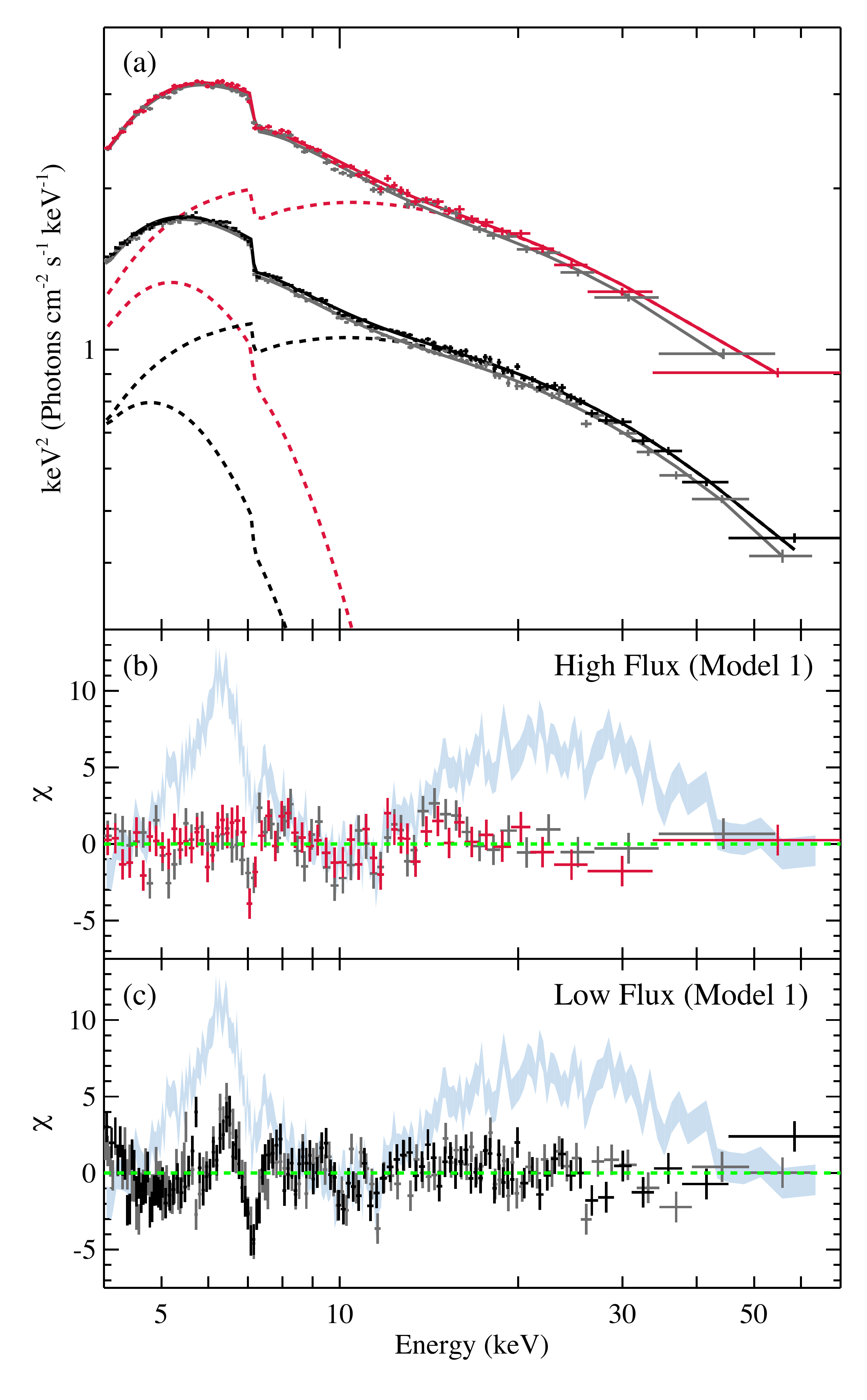}
\caption{(a) Folded \nustar\ spectra of \swiftsource\ in the intermediate state. FPMA data in the high- and low-flux epochs are plotted in red and black, respectively. FPMB data are both plotted in gray. Dashed lines mark the individual model components, a thermal disk blackbody and a cutoff power-law. (b) (c) Spectral residuals of Model 1. There is no evidence for relativistic reflection features (i.e., a broad and asymmetric Fe K$\alpha$ line and Compton reflection hump) in the data. For comparison, strong relativistic reflection features from the earlier \nustar\ observation of \swiftsource\ in the hard state reported in \cite{xu_1658} are overplotted in light blue.
\label{fig:fig5}}
\end{figure}

As shown in Figure~\ref{fig:fig5}, Model 1 fits the spectral continuum well, with the reduced chi-squared of $\chi^2/\nu=2151/1698=1.27$ ($\nu$ is the number of degrees of freedom). Allowing the parameters of {\tt $E_{\rm cut}$} and {\tt $\Gamma$} to have different values at the two epochs brings no significant improvement to the fit. The prominent spectral residuals of Model 1 are a narrow Fe K$\alpha$ emission line at $\sim$6.4--6.5~keV, and a narrow absorption feature at $\sim$7.1~keV (see Figure~\ref{fig:fig5}). The narrow emission line is only seen after the drop in flux, while the absorption line is seen in both the high- and low-flux epochs.

\capstartfalse
\begin{deluxetable}{cccc}
\tablewidth{\columnwidth}
\tablecolumns{4}
\tabletypesize{\scriptsize}
\tablecaption{Spectral Fitting Results: Part I \label{tab:tab2}}
\tablehead{
\colhead{Component}       
& \colhead{Parameter}                           
& \colhead{High-flux Epoch}
& \colhead{Low-flux Epoch}
} 
\startdata
\multicolumn{4}{c}{Model 1: TBnew*(diskbb+cutoffpl) [\nustar]} \\
\noalign{\smallskip}
\hline
\noalign{\smallskip} 
{\textsc{tbnew}}    &$N_{\rm H}$ ($\rm \times10^{23}~cm^{-2}$)  &$1.60\pm{0.04}$  &$1.48\pm{0.03}$ \\
\noalign{\smallskip}    
{\textsc{diskbb}}   &$kT_{\rm in}$ (keV)    &$1.46\pm{0.02}$    &$1.28\pm{0.01}$    \\
\noalign{\smallskip}
                    &Norm  &$74\pm5$  &$84\pm3$ \\
\noalign{\smallskip}
{\textsc{cutoffpl}} &$\Gamma$  &\multicolumn{2}{c}{$2.09\pm0.03$}  \\
\noalign{\smallskip}
                    &$E_{\rm cut}$ (keV) &\multicolumn{2}{c}{$52\pm4$} \\
\noalign{\smallskip}
                    &Norm  &$3.1\pm0.2$ &$1.7\pm0.1$ \\
\noalign{\smallskip}                    \hline                                  \noalign{\smallskip}                                                                                                          
                 & $\chi^2/{\nu}$ &\multicolumn{2}{c}{$2151/1698=1.27$}  \\
\noalign{\smallskip} 
\hline
\noalign{\smallskip} 
\multicolumn{4}{c}{Model 2: TBnew*gabs*(diskbb+cutoffpl+Gaussian) [\swift+\nustar]} \\
\noalign{\smallskip} 
\hline
\noalign{\smallskip} 
{\textsc{tbnew}}    &$N_{\rm H}$ ($\rm \times10^{23}~cm^{-2}$)  &$1.75\pm{0.03}$  &$1.41\pm{0.02}$ \\
\noalign{\smallskip}  
{\textsc{gabs}}   &$E_{\rm gabs}$ (keV)  &\multicolumn{2}{c}{$7.09\pm{0.04}$}   \\
\noalign{\smallskip} 
        &Norm  &$(1.3\pm0.4)\times10^{-2}$ &$(1.3\pm0.2)\times10^{-2}$   \\
\noalign{\smallskip} 
{\textsc{diskbb}}   &$kT_{\rm in}$ (keV)    &$1.44\pm{0.01}$    &$1.29\pm{0.01}$    \\
\noalign{\smallskip}
                    &Norm  &$86\pm4$  &$78\pm{3}$ \\
\noalign{\smallskip}
{\textsc{cutoffpl}} &$\Gamma$  &\multicolumn{2}{c}{$2.07\pm{0.03}$}  \\
\noalign{\smallskip}
                    &$E_{\rm cut}$ (keV) &\multicolumn{2}{c}{$50\pm{4}$} \\
\noalign{\smallskip}
                    &Norm  &$3.0\pm0.2$ &$1.6\pm0.1$ \\
\noalign{\smallskip}            
{\textsc{Gaussian}}   &$E_{\rm gauss}$ (keV)  &\multicolumn{2}{c}{$6.48\pm{0.04}$}   \\
\noalign{\smallskip} 
        &Norm  &$<2\times10^{-4}$      &$(9\pm1)\times10^{-4}$   \\
\noalign{\smallskip}
\hline                                      
\noalign{\smallskip}                                                                                                 
                 & $\chi^2/{\nu}$ &\multicolumn{2}{c}{$2900/2580=1.12$} \\
\noalign{\smallskip}
\hline       
\noalign{\smallskip}
\multicolumn{4}{c}{Model 3: TBnew*gabs*(diskbb+relxilllp+Gaussian) [\swift+\nustar]} \\
\noalign{\smallskip} 
\hline
\noalign{\smallskip} 
{\textsc{tbnew}}    &$N_{\rm H}$ ($\rm \times10^{23}~cm^{-2}$)  &$1.84\pm{0.03}$  &$1.55\pm{0.02}$ \\
\noalign{\smallskip}  
{\textsc{gabs}}   &$E_{\rm gabs}$ (keV)  &\multicolumn{2}{c}{$7.09\pm{0.03}$}   \\
\noalign{\smallskip} 
        &Norm  &$(2.0^{+0.5}_{-0.4})\times10^{-2}$ &$(1.8\pm0.3)\times10^{-2}$   \\
\noalign{\smallskip} 
{\textsc{diskbb}}   &$kT_{\rm in}$ (keV)    &$1.38\pm{0.02}$    &$1.17\pm{0.03}$    \\
\noalign{\smallskip}
                    &Norm  &$83^{+6}_{-8}$  &$103^{+7}_{-10}$ \\
\noalign{\smallskip}
{\textsc{relxilllp}}  &$\Gamma$  &\multicolumn{2}{c}{$2.32^{+0.04}_{-0.03}$}  \\
\noalign{\smallskip}
                         &$E_{\rm cut}$ (keV) &\multicolumn{2}{c}{$91^{+17}_{-14}$} \\
\noalign{\smallskip}
                         &$h$ ($r_{\rm g}$) &\multicolumn{2}{c}{$18\pm{8}$} \\
\noalign{\smallskip}
                         &$a^*$ ($c$J/GM$^2$) &\multicolumn{2}{c}{$0.998~(\rm fixed)$}\\
\noalign{\smallskip}
                         &$R_{\rm in}$ ($r_{\rm g}$) &\multicolumn{2}{c}{$<6.7$} \\
\noalign{\smallskip}
                         &$A_{\rm Fe}$ (solar) &\multicolumn{2}{c}{$0.91~(\rm fixed)$}\\
\noalign{\smallskip}
                         &$\theta$ ($^\circ$) &\multicolumn{2}{c}{$64~(\rm fixed)$}\\ 
\noalign{\smallskip}
                         &log $(\xi)$  &\multicolumn{2}{c}{$3.7^{+0.4}_{-0.3}$} \\
\noalign{\smallskip}
                         &$R_{\rm ref}$ &\multicolumn{2}{c}{$0.4\pm0.3$} \\
\noalign{\smallskip}
                         &Norm ($10^{-2}$) &$6.5^{+1.3}_{-1.1}$       &$3.5\pm0.6$ \\
\noalign{\smallskip}            
{\textsc{Gaussian}}   &$E_{\rm gauss}$ (keV)  &\multicolumn{2}{c}{$6.45\pm{0.06}$}   \\
\noalign{\smallskip} 
        &Norm   &$<1\times10^{-4}$      &$(7\pm1)\times10^{-4}$   \\
\noalign{\smallskip}
\hline                                      
\noalign{\smallskip}                                                                                                 
                 & $\chi^2/{\nu}$ &\multicolumn{2}{c}{$2837/2576=1.10$} \\
\noalign{\smallskip}
\hline
\noalign{\smallskip}
\multicolumn{2}{c}{$F_{\rm disk}$ (erg cm$^{-2}$ s$^{-1}$)$^a$} &$5.66\times10^{-9}$ &$3.09\times10^{-9}$  \\
\multicolumn{2}{c}{$F_{\rm powerlaw}$ (erg cm$^{-2}$ s$^{-1}$)$^a$} &$1.19\times10^{-8}$ &$6.51\times10^{-9}$ \\
\multicolumn{2}{c}{$F_{\rm total}$ (erg cm$^{-2}$ s$^{-1}$)$^a$} &$1.75\times10^{-8}$ &$9.60\times10^{-9}$ 
\enddata  
\tablecomments{
a. Unabsorbed flux in 0.1--500~keV calculated based on the normalization of \nustar/FPMA. 
}
                                
\end{deluxetable}

\begin{figure}
\centering
\includegraphics[width=0.49\textwidth]{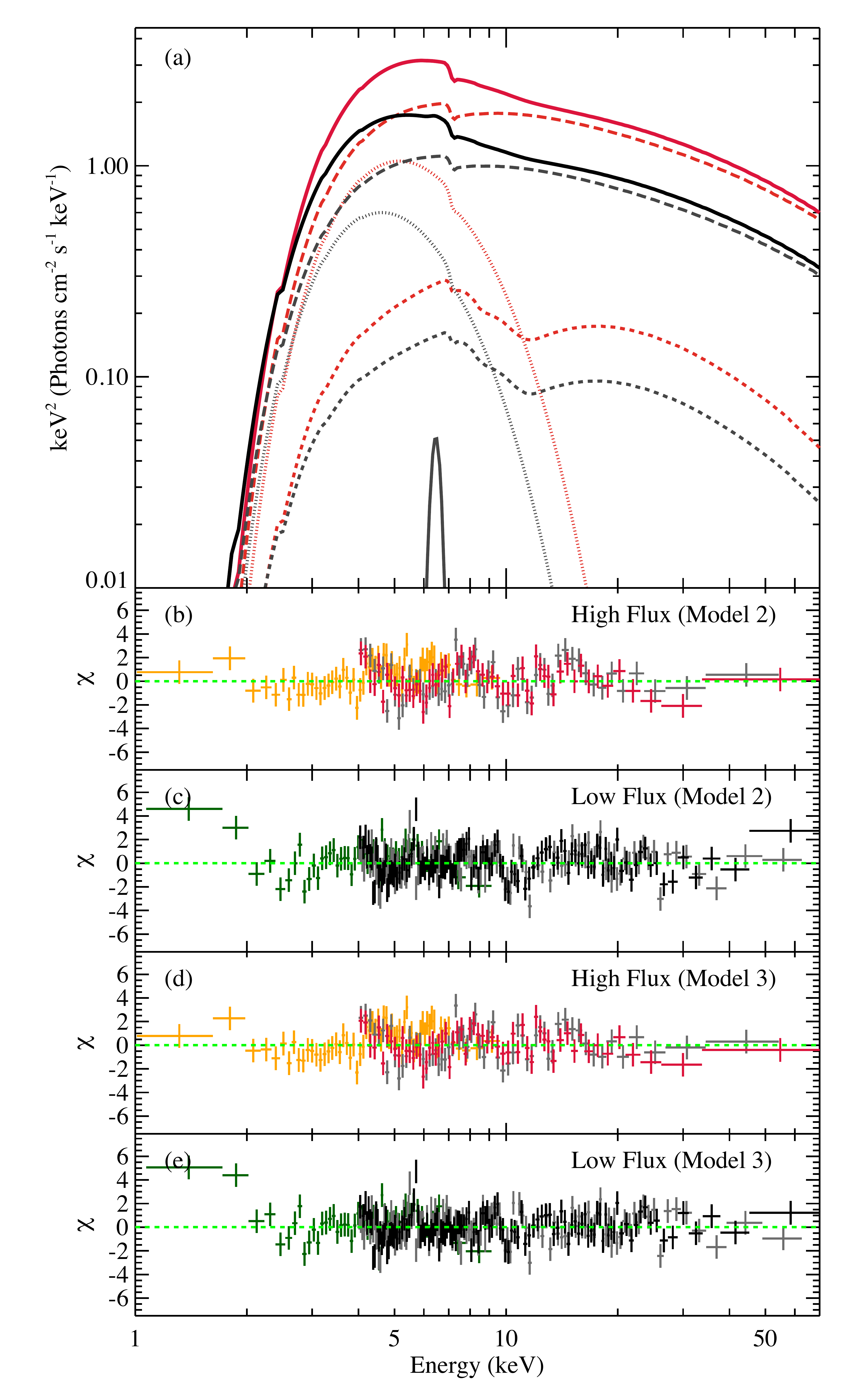}
\caption{(a) Best-fit model (Model 3) of the broadband X-ray spectra of \swiftsource, plotted in red for high-flux epoch and in black for low-flux epoch. The total model (solid lines) is plotted together with contributions from the putative weak relativistic disk reflection component (dashed lines), the coronal emission compnent (long dashed lines), the thermal disk (dotted lines) and the narrow Fe K$\alpha$ emission line (solid line). (b)--(e) Spectral residuals from Model 2 and Model 3. The \swift-XRT spectra are plotted in orange and green for the high- and the low-flux epoch, respectively. The same color scheme is used for the \nustar\ data as in previous plots.
\label{fig:fig6}}
\end{figure}

In order to achieve broadband X-ray coverage, we further include two contemporaneous \swift-XRT spectra in the spectral modeling, so that the parameters sensitive to the soft X-ray band (i.e., absorption column density, $N_{\rm H}$, and disk blackbody temperature, $T_{\rm in}$) can be better constrained. We use \nustar\ spectra in the energy range of 4--79~keV and \swift-XRT spectra in 1.0--10.0~keV, following \cite{xu_1658}. We allow the cross-normalization constants to vary freely for \nustar/FPMB and \swift-XRT, and fix the value at unity for \nustar/FPMA. To improve the fit, we add a Gaussian absorption line model, {\tt gabs} and a Gaussian emission line model, {\tt Gaussian}, to account for the spectral residuals of Model 1 shown in Figure~\ref{fig:fig5}, with the total model set up in XSPEC as: {\tt TBnew*gabs*(diskbb+cutoffpl+Gaussian)} (Model 2). For simplicity, we fix the line width, $\sigma$, of the {\tt gabs} model at 0.1~keV and the width of the {\tt Gaussian} model at 0.2~keV. During the joint spectral fitting, we link the centroid energies of {\tt gabs} and {\tt Gaussian} between epochs, while allow their corresponding strength to vary independently. The addition of these two extra model components reduces the reduced chi-squared of the fit ($\chi^2/\nu=2900/2580=1.12$), and leaves no visually evident spectral residuals (see Figure~\ref{fig:fig6}).

The shape of the broadband X-ray continuum is consistent with black hole binaries in the intermediate state \citep[][]{rem06, belloni2016}. The spectral fitting measures a soft power-law index ($\Gamma=2.07\pm0.03$) and a high disk blackbody temperature ($kT_{\rm in}\simeq1.2-1.4$~keV, the value varies between epochs), whereas the contribution of the thermal disk to the total unabsorbed flux in $0.1-500$~keV is only $\sim$32\%, indicating that \swiftsource\ was yet to enter a canonical soft (thermal dominant) state. The changes in the shape of the broadband X-ray spectrum after the large flux decrease are subtle (see Figure~\ref{fig:fig6}(a)), which are reflected by the similar values of the best-fit parameters measured for the two epochs (see Table~\ref{tab:tab2}). The values of the inner disk temperature, $kT_{\rm in}$, are measured to be $1.44\pm0.01$~keV and $1.29\pm0.01$~keV for the high- and the low-flux epoch, respectively. The absorption column density, $N_{\rm H}$, slight decreases from $(1.75\pm0.03)\times10^{23}$~cm$^{-2}$ to $(1.41\pm0.02)\times10^{23}$~cm$^{-2}$ after the flux drop. The variation in hardness ratios described in Section \ref{sec:sec3} is caused by this simultaneous decrease in the inner disk temperature and the absorption column density. Via modeling the broadband X-ray spectra, we can confidently rule out increased photoelectric absorption as the origin of the large decrease in X-ray flux.

One key difference when comparing the \nustar\ spectra of \swiftsource\ in the intermediate state with that from an earlier observation in the hard state \citep[][]{xu_1658}, besides spectral softening, is the disappearance of strong relativistic reflection features (see Figure~\ref{fig:fig5} for a comparison). Strong relativistic reflection features, including a broad and asymmetric Fe K$\alpha$ line and Compton reflection hump, are commonly found in bright black hole binaries during intermediate and soft states, enabling a measurement of the black hole spin in those systems \citep[e.g.,][]{tomsick14,walton16,miller18}. In the case of \swiftsource, although relativistic disk reflection features were clearly detected in the bright hard state in \cite{xu_1658}, there is no clear indication for such a component on top of the Comptonization continuum in the intermediate state. Considering that the flux of the hard X-ray coronal emission in the intermediate state is comparable to that in the hard state, reflection features should be easily detectable in the intermediate state as long as the reflection fraction is comparably high. Therefore, the non-detection of relativistic reflection features in this work is because of the intrinsic weakness of the relativistic reflection component rather than the source being too faint.

Lacking prominent relativistic reflection features in the data, we cannot obtain a robust measurement of the inner accretion disk parameters via reflection spectral modeling. Thus, we simply calculate the upper limit of the strength of any reflection component that comes from the inner disk.
To do so, we add a relativistic disk reflection component modeled by the {\tt relxilllp} model \citep[{\tt relxill} v1.0.2;][]{relxilla,relxillb} to the fit, with the total model set up in XSPEC as: {\tt TBnew*gabs*(diskbb+relxilllp+Gaussian)} (Model 3). The {\tt relxilllp} model assumes a lamp-post geometry for the corona, which parameterizes the disk emissivity profile by the height of the corona, $h$, above the accretion disk. A cutoff power-law continuum is included in the {\tt relxilllp} model, and is used as input for the reflection spectrum. To reduce the number of free parameters, we fix the black hole spin, $a^*=0.998$, the inclination of the accretion disk, $\theta=64^{\circ}$, and the iron abundance in the disk, $A_{\rm Fe}=0.91$, according to the best-fit values found in \cite{xu_1658}, as $a^*$ and $A_{\rm Fe}$ do not change between states and the $\theta$ is also usually assumed to remain constant. The best-fit reflection fraction measured by Model 3 is low, $R_{\rm ref}=0.4\pm0.3$, confirming the lack of a strong relativistic reflection component in the spectra (see also in Figure~\ref{fig:fig6}(a), the contribution of the putative relativistic reflection component is weak when compared to the total spectra).

We note that although the addition of a relativistically blurred reflection component improves the fit slightly with four extra parameters by $\Delta\chi^2/\Delta\nu=63/-4$, we do not consider it as significantly detected. As shown in Figure~\ref{fig:fig6}, the addition of a {\tt relxill} component in Model 3 only slightly reduces the residuals between 10~keV and 20~keV, which are likely to come from small spectral curvature not perfectly accounted for in Model 2, rather than being related to relativistic reflection features. We stress that when fitting spectra with the {\tt relxill} model, the constraint on the inner disk radius can only be confidently achieved in cases where the profile of the broad Fe K$\alpha$ line is clearly visible. Since there is no broad Fe K$\alpha$ line present in the spectral residuals in Figure~\ref{fig:fig5}, the small inner disk radius, $R_{\rm in}$, measured (see Table~\ref{tab:tab2}) could be driven by the lack of prominent blurred reflection features in the spectra. Thus the model tends to artificially make the line broad to mimic part of the continuum in the Fe K$\alpha$ band. Therefore, we consider the best-fit reflection fraction obtained in Model 3 as a crude upper limit, and do not discuss physical implications of the rest of the reflection parameters to avoid over-interpretation of the data.

The high S/N of the \nustar\ spectra of \swiftsource\ used in this work allows a comparison of the strength of secondary spectral features in the two epochs: the strength of the narrow absorption line at $\sim$$7.1$~keV modeled by {\tt gabs} can be considered as constant within errors; whereas the absolute flux of the narrow Fe K$\alpha$ emission line at $6.4-6.5$~keV modeled by {\tt Gaussian} clearly increases after the flux drop (the emission line is not detected in the high-flux epoch, see best-fit parameters in Table~\ref{tab:tab2}). Both features were also observed in \swiftsource\ with \nustar\ early on during its 2018 outburst in the hard state, and were discussed in detail in \cite{xu_1658}.

Absorption lines in the Fe K band of the X-ray spectrum of black hole binaries are commonly associated with blueshifted absorption lines from highly ionized iron (e.g., Fe {\small XXV} and Fe {\small XXVI}) generated in an accretion disk wind, which are frequently found when the accretion rate is high in the soft state \citep[e.g.,][]{miller08,ponti12,miller16_wind}. The constant strength we measured of the narrow absorption line implies an invariant disk wind component during the high- and low-flux epochs.

Narrow Fe K$\alpha$ emission lines in the spectra of black hole binaries are believed to be produced by distant reprocessing of the hard X-ray photons from the corona, which may arise from distant reflection of the coronal emission by the outer edge of a flared accretion disk, or re-emission in a disk wind. We can get an estimate of the location of the narrow Fe K$\alpha$ line emission in terms of the distance from the black hole based on the Fe K$\alpha$ line width. We replace the simple {\tt Gaussian} emission line model in Model 3 with a second {\tt relxilllp} model component (by setting ${\rm relf\_frac}=-1$, only the reflected part is used). As the line energy is consistent with being neutral, we fix the ionization parameter\footnote{The ionization parameter, $\xi$, is defined as $\xi=4\pi F_{\rm x}/n$, where $F_{\rm x}$ is the ionizing flux, and $n$ is the gas density.}, log($\xi$), of this {\tt relxilllp} model component at 0, allow the parameter of the inner disk radius, $R_{\rm in}$, to vary freely, and link all other parameters with those of the relativistic disk reflection component. We can get a lower limit of the inner radius of this {\tt relxilllp} component to be 406 $r_{\rm g}$ (where $r_{\rm g}\equiv {\rm GM}/c^2$, is the gravitational radius), which we use to represent the disk radius responsible for the narrow Fe K$\alpha$ line.  The emergence of a narrow Fe K$\alpha$ emission line in the low-flux epoch indicates the appearance of reprocessing material at a large distance from the central black hole after the flux drop. 

\section{DISCUSSION}
We have presented analyses of the coordinated \nustar\ and \xmm\ observations with  contemporaneous \swift-XRT data of the new black hole binary candidate \swiftsource, which caught the source in the intermediate state during its 2018 outburst. A rapid decrease in the source flux is observed by both telescopes, accompanied by the turn-on of a transient low-frequency QPO. The dramatic variation in flux and timing properties together with only minor changes in the broadband X-ray spectra are unusual for black hole X-ray binaries, and the physical driver of the event is uncertain.  We discuss possible causes of the uncommon properties observed in \swiftsource\ based on results from our X-ray spectral and timing analyses.

\label{sec:sec5}
\subsection{Invariance of Coronal Properties}
The power-law component extending to high energies in the X-ray spectra of black hole binaries is believed to originate from the so-called corona in the vicinity of black holes. Hot electrons in the corona up-scatter soft disk photons into the hard X-ray band. The parameters characterizing the spectral shape of the coronal emission is the photon-index, $\Gamma$, and exponential cutoff at the high energy end, $E_{\rm cut}$, which are associated with physical properties of the corona, its optical depth and the electron temperature \citep{lightman87, pet01}. 

In order to measure the physical properties of the corona, we replace the {\tt cutoffpl} component in Model 2 with the Comptonization model, {\tt compPS} \citep[][]{compps}. We assume a thermal electron distribution in the {\tt compPS} model, and perform spectral fitting assuming a slab and a spherical geometry for the corona. We link the seed photon temperature in the {\tt compPS} model with the disk blackbody temperature, $kT_{\rm in}$, in the {\tt diskbb} model. The physical model, {\tt compPS}, fits the data equally well as the phenomenological model, {\tt cutoffpl}. We cannot distinguish between the two coronal geometries based on the spectral modeling. The optical depth, $\tau$, and electron temperature, $kT_{\rm e}$, of the corona can be well constrained (see Table~\ref{tab:tab3} for the best-fit parameters). We find $\tau=1.6\pm0.1$ and $kT_{\rm e}=24^{+2}_{-1}$~keV assuming a slab geometry, and $\tau=2.1\pm0.2$ and $kT_{\rm e}=25\pm2$~keV assuming a spherical geometry, which is similar to the typical values reported in other black hole X-ray binaries \citep[e.g.,][]{santo13,sanch17}. Allowing $\tau$ and $kT_{\rm e}$ to have different values for the two epochs does not bring significant improvement to the fit, indicating that these parameters which define the coronal properties are invariant in spite of the large flux variation.

\capstartfalse
\begin{deluxetable}{cccc}
\tablewidth{\columnwidth}
\tablecolumns{4}
\tabletypesize{\scriptsize}
\tablecaption{Spectral Fitting Results: Part II \label{tab:tab3}}
\tablehead{
\colhead{Component}       
& \colhead{Parameter}                           
& \colhead{High-flux Epoch}
& \colhead{Low-flux Epoch}
} 
\startdata
\multicolumn{4}{c}{Slab corona geometry: TBnew*gabs*(diskbb+compPS+Gaussian)} \\
\noalign{\smallskip}
\hline
\noalign{\smallskip} 
{\textsc{tbnew}}    &$N_{\rm H}$ ($\rm \times10^{23}~cm^{-2}$)  &$1.64\pm{0.02}$  &$1.33\pm{0.02}$ \\
\noalign{\smallskip}
{\textsc{gabs}}   &$E_{\rm gabs}$ (keV)  &\multicolumn{2}{c}{$7.11\pm{0.04}$}   \\
\noalign{\smallskip} 
        &Norm  &$(1.3\pm0.4)\times10^{-2}$ &$(1.2\pm0.2)\times10^{-2}$   \\
\noalign{\smallskip}    
{\textsc{diskbb}}   &$kT_{\rm in}$ (keV)    &$1.38\pm{0.01}$    &$1.23\pm{0.01}$    \\
\noalign{\smallskip}
                    &Norm  &$108\pm5$  &$96\pm3$ \\
\noalign{\smallskip}
{\textsc{compPS}} &$\tau$  &\multicolumn{2}{c}{$1.6\pm0.1$}  \\
\noalign{\smallskip}
                    &$kT_{\rm e}$ (keV) &\multicolumn{2}{c}{$24^{+2}_{-1}$} \\
\noalign{\smallskip}
                    &Norm  &$169^{+16}_{-9}$ &$150^{+10}_{-8}$ \\
\noalign{\smallskip}
{\textsc{Gaussian}}   &$E_{\rm gauss}$ (keV)  &\multicolumn{2}{c}{$6.48\pm{0.04}$}   \\
\noalign{\smallskip} 
        &Norm  &$<3\times10^{-4}$      &$(1.0\pm0.1)\times10^{-3}$   \\
\noalign{\smallskip}                    \hline                                  \noalign{\smallskip}                                            & $\chi^2/{\nu}$ &\multicolumn{2}{c}{$2901/2580=1.12$}  \\
\noalign{\smallskip} 
\hline
\noalign{\smallskip} 
\multicolumn{4}{c}{Spherical corona geometry: TBnew*gabs*(diskbb+compPS+Gaussian)} \\
\noalign{\smallskip}
\hline
\noalign{\smallskip} 
{\textsc{tbnew}}    &$N_{\rm H}$ ($\rm \times10^{23}~cm^{-2}$)  &$1.63\pm{0.02}$  &$1.33\pm{0.02}$ \\
\noalign{\smallskip}
{\textsc{gabs}}   &$E_{\rm gabs}$ (keV)  &\multicolumn{2}{c}{$7.11\pm{0.04}$}   \\
\noalign{\smallskip} 
        &Norm  &$(1.3\pm0.4)\times10^{-2}$ &$(1.3^{+0.2}_{-0.1})\times10^{-2}$   \\
\noalign{\smallskip}    
{\textsc{diskbb}}   &$kT_{\rm in}$ (keV)    &$1.38\pm{0.01}$    &$1.24\pm{0.01}$    \\
\noalign{\smallskip}
                    &Norm  &$110\pm5$  &$98\pm3$ \\
\noalign{\smallskip}
{\textsc{compPS}} &$\tau$  &\multicolumn{2}{c}{$2.1\pm0.2$}  \\
\noalign{\smallskip}
                    &$kT_{\rm e}$ (keV) &\multicolumn{2}{c}{$25\pm2$} \\
\noalign{\smallskip}
                    &Norm  &$60\pm5$ &$54^{+5}_{-4}$ \\
\noalign{\smallskip}
{\textsc{Gaussian}}   &$E_{\rm gauss}$ (keV)  &\multicolumn{2}{c}{$6.48^{+0.04}_{-0.02}$}   \\
\noalign{\smallskip} 
        &Norm  &$<3\times10^{-4}$      &$(1.0\pm0.1)\times10^{-3}$   \\
 \noalign{\smallskip}        
 \hline                                 
 \noalign{\smallskip}             
                 & $\chi^2/{\nu}$ &\multicolumn{2}{c}{$2899/2580=1.12$}  \\
\noalign{\smallskip}
                                
\end{deluxetable}

The corona has also been proposed to be associated with the base of a jet \citep[e.g.,][]{markoff01,markoff05}. There is evidence that transient QPOs and fast changes in X-ray timing properties of black hole binaries are related to jet ejection activity observed in the radio band \citep[e.g.,][]{fender09, mj12}, which could be a potential physical explanation for the transient low-frequency QPO observed in \swiftsource. Radio flaring was indeed detected by ATCA around  February 24, one day before our joint \nustar\ and \xmm\ observations, when \swiftsource\ was in the same state. The source continued to be detected by subsequent radio observations in February and March (T. Russell, private communication). However, we note that associating the driver of the changes in X-ray flux and timing properties with dynamical activity of the jet or corona is hard to be reconciled with the invariant physical properties of the corona, i.e., its optical depth and electron temperature.

The turnover at the high energy end of the X-ray spectra of \swiftsource\ enables a good measurement of $E_{\rm cut}$ within the \nustar\ band. Studies of the evolution of cutoff energy during black hole X-ray binary outbursts reveal that the spectral turnover at high energies usually disappears after the the sources make the transition to the soft state \citep[e.g.,][]{joinet08, motta09}. The measurement of $E_{\rm cut}$ further supports that \swiftsource\ is yet to enter a canonical soft state during the time of our observations. We note that the spectral shape resembles the very high state (more generally classified as intermediate state here) reported in some black hole binaries \citep[named the steep power-law state in the review by][]{mcc06}, considering that the power-law and the disk component are both strong and the spectrum is dominated by the steep power-law component. Therefore, \swiftsource\ could be accreting at a high Eddington rate during the time of the observations, although the exact number is uncertain as the black hole mass and the distance are currently unknown.

\subsection{Weakness of Relativistic Disk Reflection}
Relativistic disk reflection, arsing from the innermost edge of the accretion disk, is common in the X-ray spectra of bright black hole binaries observed by \nustar\ during recent years \citep[e.g.,][]{tomsick14, furst15, walton17, xu_maxi18}, and has been observed in several sources in their very high/intermediate states \citep[e.g.,][]{king14, parker16}. We note that although we detected a narrow Fe K$\alpha$ line in this work, it is probably produced by reflection from distant material, thus is not an indicator for the innermost edge of the accretion disk. The presence of a distant reflection component in \swiftsource\ is demonstrated by the detection of a narrow Fe K$\alpha$ core on top of the broad Fe K$\alpha$ line profile by \nustar\ during the hard state \citep{xu_1658}, indicating that there are two separate reflection zones in the system. However, contrary to the earlier observation in the hard state, there is no unambiguous relativistically blurred disk reflection component detected in this work, when \swiftsource\ is in the intermediate state. The reflection fraction from the inner accretion disk we estimated here, $R_{\rm ref, intermediate}=0.4\pm0.3$, is significantly lower than the value measured in the bright hard state 8 days earlier ($R_{\rm ref, hard}=3.25$, \cite{xu_1658}). The significant decrease in relativistic reflection strength during the rising phase of a black hole binary outburst is unusual, as the relative strength of the reflection component has been known to scale positively with the X-ray spectral photon index in black hole binaries \citep[e.g.,][]{zdzi99,zdzi03,steiner16}.

Earlier detection of strong relativistic disk reflection features in \swiftsource\ led to the conclusion that the inner edge of the accretion disk reached the innermost stable circular orbit (ISCO) during its bright hard state \citep{xu_1658}. One possibility for the weakness of relativistic disk reflection in the intermediate state is that the accretion disk is truncated later on during the rising phase of the outburst, close to the time of the hard-to-soft state transition. Due to light-bending effects at the vicinity of a black hole (preferentially bending light towards the accretion disk and away from the observer), the value of the reflection fraction, $R_{\rm ref}$, should be greater than unity\footnote{The reflection fraction $R_{\rm ref}$ is defined as the ratio of the coronal intensity illuminating the disk to that reaching the observer in the {\tt relxill} model \citep{dauser16}.}, assuming that the accretion disk extends down to the ISCO. The low reflection fraction from the inner disk measured in this work could be most straightforwardly explained by a truncated accretion disk, so that a large portion of the photons from the corona falls directly into the black hole without being reflected by an optically thick accretion disk. If true, this would challenge the canonical picture for state evolutions in black hole X-ray binaries \citep[e.g.,][]{done07}, which assumes that the accretion disc is truncated at a large radius and is replaced by an advection-dominated accretion flow (ADAF) in the hard state; as the outburst develops, the inner edge of the disk gradually moves inward, and reaches the ISCO after the source makes the transition to the soft state. 

However, attributing the apparent weakness of relativistic reflection in \swiftsource\ to a truncated accretion disk is problematic, as the scenario of an optically thin ADAF inside the truncated disk would break down when the accretion rate is high in the intermediate/very high state of black hole X-ray binaries \citep[e.g.,][]{esin97, meyer07}. The high temperature and strength of the thermal disk emission also imply that the accretion disk cannot be significantly truncated. 

Alternatively, it is possible that the strength of reflection from the inner accretion disk is reduced by the fact that the system is viewed at high inclination. Due to obscuration by the accretion disk when viewed close to the disk plane, most of the coronal emission reflected by the innermost edge of the accretion disk may not be directly visible to the observer, with only the part scattered into the line of sight being observed. This greatly reduces the apparent reflection strength from the inner disk \citep[e.g,][]{wilkins15,steiner17}. Under this scenario, explaining the significant change in the relativistic reflection fraction between the hard and the intermediate state requires invoking some change in the disk-corona geometry as the X-ray spectral state evolves, e.g., an increase in the scale-height of the accretion disk due to enhanced accretion rate. This would be a natural consequence of the puffing up of the inner disk at relatively high Eddington ratios. In this case, we note that it is likely that part of the thermal disk emission is also blocked from view.

In addition, as radio observations show jet activities around the time of our X-ray observations, an outflowing corona at relativistic speeds could also potentially weaken  relativistic reflection features, because most photons would be beamed away from the accretion disk. Only considering special relativity effects, the reflection fraction could be reduced from 1 to as low as $\sim$0.05 for a viewing angle $>60^{\circ}$, depending on the the bulk velocity of the plasma \citep{belo99}.

\subsection{Interpretation of Rapid Flux Variation}

The broadband X-ray spectra we obtained of \swiftsource\ enable a good measurement of the absorption column density before and after the flux change, which rules out increased photoelectric absorption as the cause of the rapid decrease in X-ray flux. The generation of low-frequency QPOs is believed to be tied to matter in the inner accretion flow. The association of the rapid flux variation with the emergence of a transient QPO indicates that the physical origin of the flux variation involves fundamental changes in the inner accretion flow around the central black hole, rather than being purely caused by geometrical effects, e.g., a partial eclipse of the X-ray emission region by the donor star.

The connection of the appearance of QPOs to step-function-like rapid changes in count rates is reminiscent of the `flip-flop' transitions detected in some black hole X-ray binaries in their very high states \citep[e.g.,][]{miya91,taki97, cas04,bell05}, which have been proposed to have a jet origin. However, they are phenomenologically distinct from the case being discussed here in \swiftsource, as the `flip-flop' transitions are found to be repetitive over much shorter timescales ($\sim$100--1000~s), and QPOs are usually detected during the local peaks of the flux. Rapid transitions of QPOs along with flux variations have also been observed and studied in detail in the highly variable Galactic microquasar {GRS 1915+105} \citep[e.g.,][]{belloni2000, rod02}. However, the fast transitions observed in {GRS 1915+105} are accompanied by significant changes in the spectral shape: soft X-ray flux (from the accretion disk) is anti-correlated with the the hard X-ray flux (from the corona) \citep[][]{rod02,ueda02}. Therefore, changes in electron cooling rate in the corona, driven by the variable number of thermal photons emitted from the inner disk, provides a possible physical explanation for the overall flux variation in {GRS 1915+105}. In contrast, in the case of \swiftsource, emission from the thermal disk and the power-law component both dropped by $\sim$45\% at the time of the rapid flux decrease, which indicates a different physical origin. 

In terms of the X-ray spectral analysis of \swiftsource, we note that the most salient change in spectral parameters after the flux drop is a decrease in disk temperature by $\sim$15\%. The normalization parameters of the {\tt diskbb} model for the two epochs are consistent within errors, which are proportional to $\propto{R_{\rm in}}^2{\rm cos}\theta$ by definition (where $R_{\rm in}$ is the apparent inner disk radius, and $\theta$ is the disk inclination). Therefore, the observable inner edge of the accretion disk should remain at the same radius at the two flux levels, and the reduced disk blackbody emission is caused by a decrease in disk temperature at this radius. 

As shown in the long term monitoring light curve in Figure~\ref{fig:fig1}(a), the broadband X-ray flux of \swiftsource\ recovered after our joint observations, and underwent several similar large amplitude flux fluctuations during subsequent days until its flux level stabilized around MJD 58184. In the Galactic black hole X-ray binary {GRS 1915+105}, which is known to be accreting close to the Eddington limit, similar fast transitions occurring on much shorter time scales have been proposed to be associated with the \cite{lightman74} instability. Specifically, they have been proposed to be caused by the rapid removal and replenishment of matter forming the inner part of an optically thick accretion disk \citep[][]{belloni97}. Although the constant inner disk radius estimated by our spectral modeling of \swiftsource\ clearly indicates a different picture, it is still reasonable to attribute the origin of fast changes in spectral and timing properties observed in \swiftsource\ to certain instabilities in the accretion disk. Thermal and viscous instabilities in accretion disks were discussed in the Shakura-Sunyaev $\alpha$-disk model \citep{shakura76}. Assuming $\alpha=1$, \cite{miya94} estimated the thermal and viscous timescales for a 10~$M_\odot$ black hole to be 0.01~s and 1~s at the radius of 6~$r_{\rm g}$, respectively, scaling with the disk radius as $\propto R^{\frac{3}{2}}$. Given that the count rate of \swiftsource\ dropped rapidly in 40~s as shown in Figure~\ref{fig:fig2}, we can get a rough estimate of the radius where the instability is triggered, which corresponds to a radius of $\sim$1500 $r_{\rm g}$ for thermal instability, and $\sim$70 $r_{\rm g}$ for viscous instability. Thus, in the case of \swiftsource, it is most likely that disk instability arises from a relatively large distance from the central black hole. Then this propagates inwards, causing a change in the temperature of inner accretion disk in the X-ray emitting region, and triggering the QPO in the inner accretion flow.

Therefore, we propose accretion disk instability stimulated at a large radius, which propagates to affect the inner disk, as a coherent explanation for the various changes in X-ray spectral and timing properties along with the rapid flux variation observed in \swiftsource. It is in principle possible for the propagating instability to alter the global structure of the accretion disk. A change in the outer disk geometry, e.g., increased disk scale-height, could help account for the emergence of distant reflection features in the X-ray spectra after the large flux drop. It could also provide a possible explanation for the decrease in the observed coronal emission through occultation of coronal X-ray photons by an inflated inner disk along the disk plane, without invoking intrinsic changes in the physical properties of the corona.

\section{SUMMARY}
\label{sec:sec6}
In this work, we report a fast transition of spectral and timing properties observed in the new black hole binary candidate \swiftsource\ in the intermediate state. During the joint \nustar\ and \xmm\ observations, the source flux decreased rapidly by $\sim$45\% in 40~s, and simultaneously we observed a sudden turn-on of a low-frequency QPO. However, the rapid variation of X-ray flux and timing properties is accompanied by only subtle changes in the shape of the broadband X-ray spectrum, thus making it difficult to identify its physical origin.

The event observed in \swiftsource\ is phenomenologically new, distinct from reported behavior of previously known black hole X-ray binaries. We discuss disk instabilities triggered at a larger radius as a possible explanation for the various changes in the spectral and timing properties at the time of the rapid flux variation. In addition, strong relativistic disk reflection features previously reported in the bright hard state of \swiftsource\ in \cite{xu_1658} disappear in the intermediate state, although the hard X-ray coronal emission in the intermediate state is comparably strong, which is unusual for black hole binaries. We propose that the significant weakening in the apparent reflection from the inner disk could be related to the fact the source is viewed at high inclination, along with a possible change in the the accretion disk geometry during the hard-to-soft state transition. The appearance of a transient QPO together with the large X-ray flux variation makes \swiftsource\ interesting for the investigations of the nature of QPOs in black hole X-ray binaries.

\acknowledgments{
We thank the anonymous referee for helpful comments that improved the paper. We thank Tom Russell and James Miller-Jones for providing information about ATCA radio observations. D.J.W. acknowledges support from STFC Ernest Rutherford Fellowship. This work was supported under NASA contract No.~NNG08FD60C and made use of data from the \nustar\ mission, a project led by the California Institute of Technology, managed by the Jet Propulsion Laboratory, and funded by the National Aeronautics and Space Administration. We thank the \nustar\ Operations, Software, and Calibration teams for support with the execution and analysis of these observations. This research has made use of the \nustar\ Data Analysis Software (NuSTARDAS), jointly developed by the ASI Science Data Center (ASDC, Italy) and the California Institute of Technology (USA).}

\bibliographystyle{yahapj}
\bibliography{j1658.bib}
\end{document}